\documentclass[11pt]{article}
\usepackage{amsmath}
\usepackage{amsthm}
\date{}

\numberwithin{equation}{section}   %tie equation numbering to section

\newcommand{\bi}{\it\bf}

\renewcommand{\(}{\left(}
\renewcommand{\)}{\right)}

\renewcommand{\r}{\mbox{\bf R}}           %Real numbers
\newcommand{\e}{\varepsilon}            %Epsilon

\newcommand{\note}[1]{}                    %Deletes notes in bib.
\newcommand{\commentout}[1]{}

\theoremstyle{plain}
\newtheorem{thm}{Theorem}[section]
\newtheorem{lemma}[thm]{Lemma}
\newtheorem{prop}[thm]{Proposition}
\newtheorem{cor}[thm]{Corollary}

\theoremstyle{definition}
\newtheorem{definition}[thm]{Definition}

\newtheorem{Claim}{Claim}

\theoremstyle{remark}

\newcommand{\f}{\partial}

\renewcommand{\phi}{\varphi}

\newcommand{\cd}{,\ldots,}

\title{The Cosmological Time Function}

\author{Lars Andersson\footnotemark{$^*$} \\
Department of Mathematics \\
Royal Institute of Technology \\
S-100 44 Stockholm, Sweden \\
larsa\char'100math.kth.se \\
\and
Gregory J. Galloway \\
Department of Mathematics \\
University of Miami\\
Coral Gables, FL 33124, USA\\
galloway\char'100math.miami.edu 
\and
Ralph Howard$^\dagger$ \\
Department of Mathematics\\
University of South Carolina \\
Columbia, S.C. 29208, USA \\
howard\char'100math.sc.edu}

\begin{document}
\maketitle

\def\thefootnote{}
\footnotetext{
{\em \textup{1991} Mathematics Subject Classification.} Primary: 53C50   
Secondary: 83C75}
\footnotetext{\footnotemark{$^*$} Visiting Scientist: Max--Planck--Institut f\"ur Gravitationsphysik (Albert--Einstein--Institut) Schlaatzweg 1, D-14473 Potsdam, Germany. Supported in part by NFR, contract no. F-FU
4873-307.}
\footnotetext{\footnotemark{$^\dagger$} Supported in part by DEPSCoR grants
N00014-94-1-1163 and DAAH-04-96-1-0326}

\begin{abstract}
\noindent
Let $(M,g)$ be a time oriented Lorentzian manifold and $d$ the
Lorent\-zian distance on $M$. The function $\tau(q):=\sup_{p< q}d(p,q)$ is
the {\bi cosmological time function}\/ of $M$, where as usual $p< q$ 
means that $p$ is in the causal past of $q$.  This function is
called {\bi regular}\/ iff $\tau(q)<\infty$ for all $q$ and
also $\tau\to 0$ along every past inextendible causal curve.  If the
cosmological time function $\tau$ of a space time $(M,g)$ is regular
it has several pleasant consequences: (1)~It forces $(M,g)$ to be
globally hyperbolic, (2)~every point of $(M,g)$ can be connected to
the initial singularity by a rest curve (i.e., a timelike geodesic
ray that maximizes the distance to the singularity), (3)~the function
$\tau$ is a time function in the usual sense, in particular (4)~$\tau$ is
continuous, in fact locally Lipschitz and the second
derivatives of $\tau$ exist almost everywhere.
\end{abstract}

\thispagestyle{empty}
%\tableofcontents

%\newpage
%\ \ \ 
%\thispagestyle{empty}

%\newpage
%\setcounter{page}{1}
\section{Introduction.}

Time functions play an important role in general relativity.
They arise naturally in the global causal theory of spacetime
and they
permit a decomposition of spacetime into space and time which is useful,
for example, in the study of the solution of the Einstein equation.
The choice of a time function, however, can be rather arbitrary and
a given time function may have little
physical significance.
Very few situations have been identified which lead to a canonically
defined time function.  In this paper we introduce and study what may
be viewed in the cosmological setting as a canonical time function.

Let $(M,g)$ be a spacetime (i.e., a time oriented Lorentzian Manifold)
and  let $d:M\times M\to
[0,\infty]$ be the Lorentzian distance function.  Define the
{\bi cosmological time  function}\/ $\tau:M\to (0,\infty]$ by
\begin{equation}\label{time-def}
\tau(q):=\sup_{p < q}d(p,q)
\end{equation}
If $c$ is a causal curve in $M$ denote by $L(c)$ the Lorentzian length of
$c$ and for $q \in M$, let  ${\cal C}^-(q)$ be the set of all past directed causal
curves $c$ in $M$ that start at $q$.  Then we have the alternative definition
$$
\tau(q):=\sup\{L(c): c\in {\cal C}^-(q)\}.
$$
The number $\tau(q)$ can be thought of as the length of time the point $q$
has been in existence.

In general the function $\tau$ need not be at all nice.  For example in the
case of flat Minkowski space $\tau\equiv \infty$.  We will give examples
below where $(M,g)$ is globally hyperbolic, $\tau(q)<\infty$ for all $q$ 
but $\tau$ is discontinuous.

\begin{definition}\label{def:reg}
The cosmological time function $\tau$ of $(M,g)$ is {\bi regular}\/ if and
only if
\begin{enumerate}
\item \label{point:finite}
$(M,g)$ has {\bi finite existence times}, i.e., $\tau(q)<\infty$
for all $q\in M$.
\item \label{point:tau0}
$\tau\to0$ along every past inextendible causal curve.
\end{enumerate}
\end{definition}

The first of these conditions is an assertion that the spacetime has an
initial singularity in the strong sense that for each point of the
spacetime any particle that passes through $q$ has been in existence for
at most a time of $\tau(q)$.  The second condition is a weak
completeness assumption.  It asserts that if we believe that the condition
$\tau=0$ defines the initial singularity and that world lines of particles
are inextendible, then every particle came into existence at the initial
singularity.

Our main result is that if the cosmological time
function is regular then the spacetime is quite well behaved.

\begin{thm} \label{thm:timereg}
Suppose $(M,g)$ is a spacetime such that the function $\tau:M\to
(0,\infty)$ defined
by (\ref{time-def})
is regular.  Then the following properties hold.
\begin{enumerate}
\item $(M,g)$ is globally hyperbolic.
\item $\tau$ is a time function in the usual sense, i.e., $\tau$ is
continuous and is strictly increasing along future directed causal curves.
\item For each $q\in M$ there is a future directed timelike ray
$\gamma_q:(0,\tau(q)]\to M$
that realizes the distance from the ``initial singularity'' to $q$, that is,
$\gamma_q$ is a future directed timelike unit speed geodesic, which is
maximal on each segment,
such that,
\begin{equation}
\gamma_q(\tau(q))=q,\qquad \tau(\gamma_q(t))=t,\quad \mbox{for $t\in
(0,\tau(q)]$}.
\end{equation}
\item The tangent vectors  $\{\gamma_q'(\tau(q)):q\in M\}$ are locally
bounded away from the light
cones.  More precisely, if $K\subseteq M$ is compact then
$\{\gamma_q'(\tau(q)):q\in K\}$ is
a bounded subset of the tangent bundle $T(M)$.
\item $\tau$ has the following additional regularity property: it is locally
Lipschitz and
its first and second
derivatives exist almost everywhere.
\end{enumerate}
\end{thm}

Conditions similar to Property 4 have played an important role in the
analysis of the
regularity of Lorentzian Busemann functions
and their level sets (cf., \cite{AGH-max}, \cite{Galloway-Horta}).  Here
Property 4 will be used to establish Property~5.

  For regularity properties of the level sets $\{\tau =a\}$ see
Section~\ref{sec:level} (as well as the corollary at the end of
Section~\ref{sec:proofs}).
The various conclusions of the
theorem will be proven as separate propositions in the following sections.

\subsection{Terminology and notation.}  We use the standard
terminology and notation from Lorentzian geometry, following for
example~\cite{Hawking-Ellis}.  In particular if $(M,g)$ is a spacetime
then $p\ll q$ (respectively, $p<q$) means there is a future directed timelike
(resp. causal) curve from $p$
to $q$.  If $S\subset M$ then $I^+(S)$ is the chronological future of
$S$ and $J^+(S)$ is the causal future of $S$.  Likewise $I^-(S)$ and
$J^-(S)$ are the chronological past and causal past of $S$.  If
$p < q$, then the Lorentzian distance $d(p,q)$ is the supremum of the
lengths of all the future directed causal curves from $p$ to $q$ and
if $p\not < q$ then $d(p,q)=0$.  A fact that will be used repeatedly is
that if $x < p < q$ then the {\bi reverse triangle inequality}\/
$$
d(x,q)\ge d(x,p)+d(p,q)
$$
holds.

\section{Proofs of the basic properties of the cosmological time function.}
\label{sec:proofs}
\subsection{Continuity of the cosmological time function.}

\begin{prop}\label{time-cont}
If the cosmological time function $\tau$ of $(M,g)$ is regular then it
is continuous and satisfies the reverse Lipschitz inequality
\begin{equation}\label{reverse-Lip}
p < q\quad \mbox{implies}\quad \tau(p)+d(p,q)\le \tau(q).
\end{equation}
\end{prop}

\begin{proof} For any $p\in M$ the function $q\mapsto d(p,q)$ is lower
semicontinuous   on $M$.  (That is $\liminf_{x\to q}d(p,x)\ge d(p,q)$).
For example cf.~\cite[p215]{Hawking-Ellis}. Then
$\tau(q)=\sup_{p< q}d(p,q)$
is a supremum of lower semicontinuous functions and therefore also
lower semicontinuous.  Thus to prove  continuity of $\tau$ it
is enough to show it is  upper-semicontinuous, that is
$\limsup_{x\to q}\tau(x)\le \tau(q)$.

Assume, toward a contradiction, that $\tau$ is not upper semicontinuous
at $q\in M$.  Then there is $\e>0$ and a sequence $x_\ell\to q$ such that
for each $\ell$
$$
\tau(x_\ell)\ge \tau(q)+\e.
$$
For each $\ell$ we can choose $p_\ell$ with
$$
d(p_\ell,x_\ell)\ge \tau(x_\ell)-\frac1\ell.
$$
Moreover, by the regularity of $\tau$, we can choose the sequence
$\{p_\ell\}$ so that $\tau(p_\ell)\to 0$ as $\ell\to0$. (To see this make
any choice of $\hat{p}_\ell$ with $d(\hat{p}_\ell,x_\ell)\ge
\tau(x_\ell)-1/\ell$.  Then choose a past directed inextendible curve
$\sigma$ starting at $\hat{p}_\ell$.  By the definition of regular there is a
point $p_\ell$ on $\sigma$ with $\tau(p_\ell)<1/\ell$.  Then
$d(p_\ell,x_\ell)\ge d(\hat{p}_\ell,x_\ell)\ge \tau(x_\ell)-1/\ell$ and
$\lim_{\ell\to\infty}\tau(p_\ell)=0$.)  The condition $\tau(p_\ell)\to0$
and the lower
semicontinuity of $\tau$ implies that $\{p_\ell\}$ diverges to infinity,
that is it has no convergent subsequences.

We now put a complete Riemannian metric $h$ on $M$ and assume that all
causal curves (except possibly those arising as limit curves) are
parameterized with respect to arc length in the metric~$h$.  Since
$d(p_\ell,x_\ell)<\infty$ there is a past directed causal curve
$c_\ell:[0,a_\ell]\to M$ (parameterized with respect to arc length in $h$)
from $x_\ell$ to $p_\ell$ such that
\begin{equation}\label{L-bd}
L(c_\ell)\ge d(p_\ell,x_\ell)-\frac1\ell
        \ge\tau(x_\ell)-\frac2\ell
        \ge\tau(q)+\e-\frac2\ell
\end{equation}
where $L(\cdot)$ is the Lorentzian arc length functional.
Since $\{p_\ell\}$ diverges, $a_\ell\to\infty$.  Hence, by passing to a
subsequence if necessary, we have that $\{c_\ell\}$ converges uniformly on
compact sets to a past inextendible timelike or null ray (maximal half
geodesic) $c:[0,\infty)\to M$ (cf.~\cite[Sections~2--3]{Eschenburg-Galloway}).
Moreover, by the upper semicontinuity of the Lorentzian arclength
functional (strong causality is not required, again
cf.~\cite{Eschenburg-Galloway}), for each $b>0$
\begin{equation}\label{eq:up-len}
L(c\big|_{[0,b]})\ge \limsup_{\ell\to\infty}L(c_{\ell}\big|_{[0,b]}).
\end{equation}

\begin{Claim}\label{claim1} The curve $c:[0,\infty)\to M$ is null.\end{Claim}

If not then $c$ is a timelike ray.  Choose $t>0$ and
$\delta>0$ so that
$$
L(c\big|_{[0,t]})+\delta\le\frac{\e}2.
$$
By~(\ref{eq:up-len}) there is an $N$ such that for all $\ell\ge N$,
$$
L(c_\ell\big|_{[0,t]})\le L(c\big|_{[0,t]})+\delta \le \frac{\e}2.
$$
Hence, by~(\ref{L-bd}) and the above,
$$
L(c_\ell\big|_{[t,a_\ell]})=L(c_\ell)-L(c_\ell\big|_{[0,t]})
        \ge \tau(q)+\frac{\e}2 - \frac2\ell.
$$
Thus when $\ell$ is sufficiently large,
$$
L(c_\ell\big|_{[t,a_\ell]})>\tau(q).
$$
On the other hand, since $c$ is timelike, we have that $c_\ell(t)\in
I^-(q)$ for all $\ell$ sufficiently large.  It follows that
$\tau(q)\ge L(c_\ell\big|_{[t,a_\ell]})>\tau(q)$.  This contradiction
establishes
the claim.
\smallskip

\begin{Claim}\label{claim2} For each $y$ on $c$, $y\ne q$ and each
neighborhood $U$ of $y$ there is a $\overline{y}\in U$ so that
$\tau(\overline{y})\ge \tau(y)+\e/2$.
\end{Claim}

We have $y=c(b)$ for some $b>0$.  Since $c$ is null~(\ref{eq:up-len})
implies
$$
L(c_\ell\big|_{[0,b]})\to 0.
$$
Let $y_\ell:=c_\ell(b)$.  Then
\begin{align*}
\tau(y_\ell)&\ge L(c_\ell\big|_{[b,a_\ell]}) =
L(c_\ell)-L(c_{\ell}\big|_{[0,b]})\\
        &\ge \tau(y)+\e-\frac2\ell-L(c_\ell\big|_{[0,b]})
        \ge \tau(y)+\frac\e2
\end{align*}
for all sufficiently large $\ell$.  Therefore given the neighborhood $U$ of
$y$ we see that the claim holds with $\overline{y}=y_\ell$ for some
sufficiently large $\ell$.
\smallskip

We now use Claim~\ref{claim2} to construct a past inextendible causal curve
along which $\tau$ does not go to zero:
\begin{enumerate}
\item First choose $y_\ell=c(b_\ell)$, $b_\ell\to\infty$ so that
$\lim_{\ell\to\infty}y_\ell$ does not exist.
%(That is $\{y_\ell\}$ diverges to infinity.)
\item Then choose $\{z_\ell\}\subset M$ so that
\begin{enumerate}
\item $z_{\ell+1}\in I^-(z_\ell)$,
\item $z_\ell\in I^+(y_\ell)$,
\item $\lim_{\ell\to\infty}z_\ell$ does not exist.
\end{enumerate}
\end{enumerate}
Let $\tilde{c}$ be a past directed timelike curve which threads through
$z_1\gg  z_2\gg  z_3\gg  \cdots$.
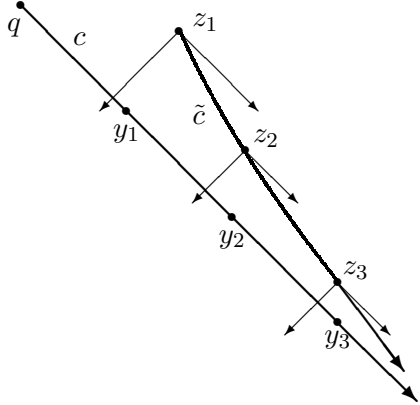
\begin{figure}[ht]
\begin{center}
\begin{picture}(150,150)

\put(0,150){\circle*{3}}
\put(40,110){\circle*{3}}
\put(80,70){\circle*{3}}
\put(120,30){\circle*{3}}
\put(-5,140){$q$}
\put(35,100){$y_1$}
\put(75,60){$y_2$}
\put(115,22){$y_3$}

\put(60,140){\circle*{3}}
\put(85,95){\circle*{3}}
\put(120,45){\circle*{3}}

\put(65,142){$z_1$}
\put(88,98){$z_2$}
\put(122,48){$z_3$}

\put(20,135){$c$}
\put(65,105){$\tilde{c}$}

\thicklines
\put(0,150){\vector(1,-1){150}}
\bezier{300}(60,140)(80,95)(120,45)
\put(120,45){\vector(3,-4){25}}

\thinlines
\put(60,140){\vector(1,-1){30}}
\put(60,140){\vector(-1,-1){30}}
\put(85,95){\vector(1,-1){20}}
\put(85,95){\vector(-1,-1){20}}
\put(120,45){\vector(1,-1){20}}
\put(120,45){\vector(-1,-1){20}}

\end{picture}
\end{center}
\end{figure}

As $\lim_{\ell\to\infty}z_\ell$ does not exist, the
curve $\tilde{c}$ is past inextendible.  Also $I^-(z_\ell)$ is a
neighborhood of $y_\ell$ therefore by Claim~\ref{claim2} there is a
$\overline{y}_\ell$ in $I^-(z_\ell)$ with $\tau(\overline{y}_\ell)\ge
\tau(q)+\e/2$.  But for any point $p$ on $c$ there is a $\ell$ with
$z_\ell\ll p$ and therefore $\tau(p)\ge \tau(z_\ell)\ge \tau(q)+\e/2$.
As $p$ was any point of $\overline{c}$ this contradicts that $\tau\to0$
along every past inextendible causal curve and proves the continuity of $\tau$.

To prove the reverse Lipschitz inequality assume $p < q$ and let $x < p$.
Then $x < q$ and so by the reverse triangle inequality
(i.e. $d(x,p)+d(p,q)\le d(x,q)$),
$$
\tau(p)+d(p,q)=\sup_{x < p}(d(x,p)+d(p,q))\le \sup_{x < p}d(x,q)
\le \sup_{x < q}d(x,q)=\tau(q).
$$
This completes the proof.~\end{proof}

%\begin{prop}\label{time} If the cosmological time function $\tau$ is
%regular, then it is a time function.  That is $\tau$
%is continuous and strictly increasing along future directed casual
%curves.
%\end{prop}
%
%
%\begin{proof}
%We already know that $\tau$ is continuous.  Let $\sigma:(a,b)\to M$ be a
%future directed causal curve and let $t_1,t_2\in (a,b)$ with $t_1<t_2$.
%Set $p=\sigma(t_1)$ and $q=\sigma(t_2)$.  Then we need to show
%$\tau(p)<\tau(q)$.  If
%\end{proof}

\subsection{Global hyperbolicity of $(M,g)$}

\begin{prop}\label{global-hyp}
Let $(M,g)$ be a spacetime so that the cosmological time function $\tau$
is regular. Then $(M,g)$ is globally hyperbolic.
\end{prop}

\begin{proof} We have shown in Proposition~\ref{time-cont} that $\tau$ is
continuous.
Therefore, if $S_t:=\{q\in M: \tau(q)=t\}$ then by elementary
topological and causal considerations, $S_t$ is closed,
achronal, and edge-less.  (That $S_t$ is achronal follows from
the reverse Lipschitz inequality.  That $S_t$ is closed and edge-less follows
from the continuity of $\tau$.)

%(That $S_t$ is closed follows from the
%continuity of $\tau$.  Recall a subset $S$ of a spacetime is achronal
%if no two points of $S$ are timelike related.  But if $p\ne q$ are
%timelike related then, by relabeling if needed, we can assume $p\ll q$
%so that $d(p,q)\ne 0$.  Therefore the reverse Lipschitz
%inequality~(\ref{reverse-Lip}) implies $\tau(p)<\tau(q)$.  This
%implies $S_t$ is achronal.  If $S\subset M$ is an achronal set then
%the edge of $S$ is the set of points $p$ in the closure $\overline{S}$
%so that for every neighborhood $U$ of $p$ there are points $p_-\in
%I^-(p,U)$ (the chronological past of $p$ within $U$)
%and $p_+\in I^+(p,U)$ so that $p_-$ can be joined to
%$p_+$ by a timelike curve in $U$ that does not intersect $S$
%(cf.~\cite[p202]{Hawking-Ellis}).
%But then the continuity of $\tau$ implies the level sets $S_t$ are all
%edge-less.)

Recall that the future domain of dependence $D^+(S_t)$ of $S_t$ is the
set of all points $q\in M$ such that every past inextendible causal curve
from $q$  intersects $S_t$.
%Likewise the past domain of dependence
%$D^-(S_t)$ of $S_t$ is the set of $p\in M$ so that every future
%inextendible curve $c$ passing through $p$ intersects $S_t$.
The past domain of dependence $D^-(S_t)$ is defined time-dually.
The domain
of dependence of $S_t$ is $D(S_t)=D^+(S_t)\cup D^-(S_t)$. From 
the definition of regularity and the continuity of $\tau$ we see that
$D^+(S_t)=\{q:\tau(q)\ge t\}$. It follows that each point of $M$ is contained
in ${\rm int\/}\, D^+(S_t)$ for some $t$.  Since strong causality holds at
each point of
${\rm int\/}\, D^+(S_t)$ (cf.~\cite[Prop~5.22~p48]{Penrose:topology}),
$(M,g)$ is strongly causal.

Now let $p,q\in M$ with $p< q$.
Then choose $t>0$ with $t<\tau(p)$.  Then $J^-(q)\cap
J^+(p)$ is a subset of the open set $\{x:\tau(x)>t\}\subset D(S_t)$ and
thus $J^-(q)\cap J^+(p)$ is contained in the interior of $D(S_t)$.  This
implies (cf.~\cite[Prop~5.23~p48]{Penrose:topology}) $J^-(q)\cap J^+(p)$ is
compact.  As $(M,g)$ is strongly causal and $p$ and $q$ were arbitrary
points of $M$ with $p< q$, this verifies the definition of globally
hyperbolic.~\end{proof}

\subsection{Existence of maximizing rays to the initial singularity}

\begin{prop}\label{prop:rays}
Let $(M,g)$ be a spacetime with regular cosmological time function $\tau$.  
Then for each
$q\in M$ there is a future directed timelike
ray $\gamma_q:(0,\tau(q)]\to M$ that realizes the
distance from the ``initial singularity'' to $q$.  That is, $\gamma_q$
is a future directed timelike unit speed geodesic that
realizes the distance between any two of its points  (for \,$0<s<t\le\tau(q)$,
$d(\gamma_q(s),\gamma_q(t))=t-s$) and satisfies,
\begin{equation}\label{int-ray}
\gamma_q(\tau(q))=q,\qquad \tau(\gamma_q(t))=t,\quad \mbox{for $t\in
(0,\tau(q)]$}.
\end{equation}

\end{prop}

\begin{proof}
For the purpose of the proof we will parameterize curves with respect to a
complete Riemannian metric $h$ on $M$ as in the proof of
Proposition~\ref{time-cont}.  Fix $q\in M$.  As in the
proof of Proposition~\ref{time-cont}, one can construct a sequence
$\{y_\ell\}\subset I^-(q)$ that diverges to infinity and such that
$$
d(y_\ell,q)\ge \tau(q) -\frac{1}{\ell}\quad \text{and}\quad
\tau(y_\ell)<\frac1\ell.
$$
By Proposition~\ref{global-hyp}, $(M,g)$ is globally hyperbolic so there is
a past directed maximal geodesic segment $\gamma_\ell:[0,a_\ell]\to M$ from
$q=\gamma_\ell(0)$ to $y_\ell=\gamma_\ell(a_\ell)$.  Since $\{y_\ell\}$
diverges to infinity and the curves are parameterized with respect to
$h$-arclength we have $a_\ell\to \infty$.  Hence, by passing to a
subsequence if necessary, the sequence $\{\gamma_\ell\}$ converges to a past
inextendible timelike or null ray $\gamma:[0,\infty)\to M$ based at
$q=\gamma(0)$.  Hence for all $b\in (0,a_\ell)$,
\begin{equation}\label{A}
L(\gamma\big|_{[0,b]})=d(\gamma(b),q).
\end{equation}

\noindent
{\bf Claim.\/} $\gamma$ is timelike and for each $b\in(0,\infty)$,
\begin{equation}\label{B}
d(\gamma(b),q)=\tau(q)-\tau(\gamma(b)).
\end{equation}
Hence by suitably reparameterizing $\gamma$ we obtain a timelike ray
$\gamma_q$ that satisfies~(\ref{int-ray}).

To see the claim holds first note by the reverse Lipschitz inequality,
\begin{equation}\label{q-rev}
d(\gamma(b),q)\le \tau(q)-\tau(\gamma(b)).
\end{equation}
By the maximality of the segments $\gamma_\ell$,
$$
d(\gamma_\ell(b),q)=d(y_\ell,q)-d(y_\ell,\gamma_\ell(b))\ge
\(\tau(q)-\frac1\ell\) -\tau(\gamma_\ell(b)).
$$
Letting $\ell\to \infty$ we obtain
$d(\gamma(b),q)\ge\tau(q)-\tau(\gamma(b))$ which, together
with~(\ref{q-rev}), establishes~(\ref{B}).  Moreover since
$\tau(\gamma(b))\to 0$ as $b\to \infty$, by taking $b$ large enough
in~(\ref{B}) we see that $d(\gamma(b),q)>0$ and thus $\gamma$ must be
timelike. This completes the proof of the claim and all of the
proposition save the last statement about $\gamma_q$ realizing the
distance between its points.  But this follows easily from the reverse
Lipschitz inequality for $\tau$.~\end{proof}

\begin{prop}\label{speed-bd}
Assume the cosmological time function $\tau$ of $M$ is regular and that
$K\subset M$ is compact.  For each $q\in K$ let $\gamma_q:(0,\tau(q)]\to M$
be a maximizing ray from the initial singularity to $q$ in the sense
that~(\ref{int-ray}) holds.  Then $\{\gamma_q'(\tau(q)):q\in K\}\subset T(M)$
is bounded in $T(M)$ (or, what is the same thing, $\{\gamma_q'(\tau(q)):q\in
K\}$ has compact closure in $T(M)$).
\end{prop}

\begin{proof}
The proof is similar to the last proposition and again we parameterize
curves with respect to a complete Riemannian metric on $M$.  If
$\{\gamma_q'(\tau(q)):q\in K\}$ is not bounded then
there exist inextendible timelike rays
$\gamma_\ell:[0,\infty)\to M$, parameterized with respect to $h$-arclength,
which satisfy
\begin{equation}\label{D}
d(\gamma_\ell(b),\gamma_\ell(0))=\tau(\gamma_\ell(0))-\tau(\gamma_\ell(b))
\end{equation}
for all $b\in (0,\infty)$, such that $\gamma_\ell(0)\to q\in K$ and the
$h$-unit vectors $\gamma'_\ell(0)$ converge to an $h$-unit vector $X$ which
is null in the Lorentzian metric.  Let $\gamma:[0,\infty)\to M$ be the past
inextendible null geodesic parameterized with respect to $h$-arclength,
satisfying $\gamma(0)=q$ and $\gamma'(0)=X$.  Then $\gamma$ is necessarily
a null ray (otherwise the maximality of the $\gamma_\ell$'s would be
violated).  By (\ref{D}) we have
\begin{align*}
d(\gamma(b),\gamma(0))&=\lim_{\ell\to\infty}d(\gamma_\ell(b),\gamma_\ell(0))\\
        &=\lim_{\ell\to\infty}(\tau(\gamma_\ell(0))-\tau(\gamma_\ell(b)))
        =\tau(\gamma(0))-\tau(\gamma(b))>0
\end{align*}
for sufficiently large $b$.  But this contradicts that $\gamma$ is a null
ray.~\end{proof}

\subsection{$\tau$ is strictly monotone on causal curves.}

\begin{prop}\label{time-fcn}
If the cosmological time function $\tau$ is regular then it is  a time
function in the usual sense, that is, it is continuous and strictly
increasing along future
directed causal curves.
\end{prop}

\begin{proof} We have already shown $\tau$ is continuous.  Let
$\sigma:(a,b)\to M$ be a future directed causal curve and $t_1,t_2\in
(a,b)$ with $t_1<t_2$.  Set $p:=\sigma(t_1)$ and $q:=\sigma(t_2)$.  If
$d(p,q)>0$ then $\tau(q)\ge\tau(p)+d(p,q)>\tau(p)$ by the reverse Lipschitz
inequality for $\tau$.  Thus assume $d(p,q)=0$.  Then there is a null
geodesic ray $\eta$ from $p$ to $q$.  Let $\gamma_p$ be the timelike ray to
$p$ guaranteed by Proposition~\ref{prop:rays}.  Choose a point $x$ on
$\gamma_p$ to the past of $p$.  Then by a ``cutting the  corner''
argument near $p$ strict inequality holds in the reverse triangle
inequality.  This strict inequality and $d(p,q)=0$ imply
$$
d(x,q)>d(x,p)+d(p,q)=d(x,p).
$$
Hence,
$$
\tau(q)-\tau(p)\ge d(x,q)>d(x,p)=\tau(p)-\tau(x)
$$
which implies $\tau(p)>\tau(q)$, as desired.~\end{proof}

Recall that for a closed subset $S \subset M$, the {\bi future Cauchy
horizon} 
$H^+(S)$ is by definition future boundary of the domain of dependence
$D^+(S)$, 
$$
H^+(S) = \overline{D^+(S)} - I^-(D^+(S)).
$$
$H^-(S)$ is defined analogously.
If $S \subset M$ is edge--less and acausal, then $S$ is called a {\bi partial
Cauchy surface} and if in addition $H^+(S) = \emptyset$ then $S$ is called a
{\bi future Cauchy surface}, see \cite[Chapter 6]{Hawking-Ellis} for details. We can
now state the following corollary. 
\begin{cor}
If the cosmological time function $\tau$ is regular then the level sets
$S_a:=\{q:\tau(q)=a\}$ (if nonempty) are 
%partial Cauchy surfaces
%(that is edge-less and acausal), and are 
future Cauchy surfaces.
\end{cor}
\begin{proof}
As observed in Proposition~\ref{global-hyp}, $S_a$ is edge-less.
The acausality of $S_a$ is
immediate from Proposition~\ref{time-fcn}.  Suppose $H^+(S_a) \ne \emptyset$.
Let $\eta$ be a past
inextendible null geodesic generator of $H^+(S_a)$ with future end
point $q \in H^+(S_a)$ (cf. \cite[Prop.6.5.3 p203]{Hawking-Ellis}).
Since $q\in I^+(S_a)$, $\tau(q) > a$.  But then, since $\tau \to 0$ along
$\eta$ and $\tau$ is
continuous, there is a point $p$ on $\eta$ such that $\tau(p)=a$, i.e.,
$\eta$ meets $S_a$,
which cannot happen.
\end{proof}

Simple examples show that the level sets $S_a$ need not be Cauchy,
i.e., $H^-(S_a)$ need not be empty.

\section{Other regularity properties of $\tau$ and its level sets}
\label{sec:level}

A continuous function $u$ defined on an open subset $U$ of $\r^n$ is
{\bi semiconvex}\/ if and only if each point $x\in U$ there is a
smooth function $f$ defined near $x$ so that $u+f$ is convex in a
neighborhood of $x$.  Using Lemma~\ref{quad-spt} below it is not hard
to check that the class of semiconvex functions is closed under
diffeomorphisms between open subsets of $\r^n$ and therefore the
definition of semiconvex extends to smooth manifolds
(cf.~\cite{Bangert:gen-convex}). By a well known theorem of
Aleksandrov a convex function has first and second derivatives almost
everywhere and thus a semiconvex function has the same property.  (For
a beautiful recent proof see~\cite[Thm~A.2~p~56]{users-guide}).

\begin{prop}\label{semiconvex}
If the cosmological time function $\tau$ is regular on $(M,g)$ then it is
semiconvex and thus its first and second derivatives exist at almost
all points of $M$.
\end{prop}

If $f$ is a smooth function on an open subset of $\r^n$ then denote by
$D^2f$ the matrix of second second partial derivatives of $f$. Let $I$ be
the $n\times n$ identity matrix.  For a constant $c$ let $D^2f(x)\le cI$
mean that $cI-D^2f(x)$ is positive semidefinite.  Also recall that if $u$
is continuous then a smooth function $\phi$ is a lower support function for
$u$ at $x_0$ iff both $u$ and $\phi$ are defined in a neighborhood of
$x_0$, $u(x_0)=\phi(x_0)$ and $ \phi\le u$ near $x_0$.  The proof of the
proposition is based on the following lemma.

\begin{lemma}\label{quad-spt}
Let $U\subset \r^n$ be convex and let $u:U\to \r$ be continuous.  Assume
for some constant $c$  and all $q\in U$ that $u$ has a  lower support
function $\phi_q$ at $q$ so that $D^2\phi(x_0)\ge cI$.  Then $u-c\|x\|^2/2$
is convex in $U$ and therefore $u$ is semiconvex.
\end{lemma}

\begin{proof}
While in some circles this is a well known folk-theorem, the only explicit
reference we know is~\cite[Sec.~2]{AGH-max}.~\end{proof}

\begin{proof}{Proof of Proposition~\ref{semiconvex}}
For any point $q\in M$ let $\gamma_q:(0,\tau(q)]\to M$ be a geodesic
segment realizing the distance from the initial singularity to $q$ as
in Proposition~\ref{prop:rays}. Define a function $\phi_q$ on
$I^+(\gamma_q(\tau(q)/2))$ by
$$
\phi_q(x):=\tau(q)/2+d(\gamma_q(\tau(q)/2),x).
$$
By Proposition~\ref{prop:rays}, $\gamma_q$ realizes the distance between any
two of its points and thus $d(\gamma_q(t),q)=\tau(q)-t$ for $t\in (0,\tau(q)]$.
Hence
$$
\phi_q(q)=\tau(q)/2+d(\gamma_q(\tau(q)/2),q)=\tau(q).
$$
By the reverse Lipschitz inequality for $\tau$, if $x\in
I^+(\gamma_q(\tau(q)/2))$
$$
\tau(x)-\tau(\gamma_q(\tau(q)/2))\ge d(\gamma_q(\tau(q))/2,x),
$$
which implies $\tau(x)\ge \phi_q(q)$ and thus $\phi_q$ is a lower support
function for $\tau$ at $q$.

Also as $\gamma_q$ maximizes the distance between its points the segment
$\gamma_q\big|_{[\tau(q)/2,\tau(q)]}$ will be free of cut points.  Thus the
map $x\mapsto d(\gamma_q(\tau(q)/2),x)$ is smooth in a neighborhood of $q$.
This implies $\phi_q$ is smooth near $q$.
By standard comparison theorems (see e.g., \cite{AH-comp,Eschenburg-comp})
it is possible to give upper and lower bounds for the Hessian (defined in
terms of the metric connection of $(M,g)$) of
$x\mapsto d(\gamma_q(\tau(q)/2),x)$ just in terms of upper and lower bounds
of the timelike sectional curvatures of two planes containing $\gamma_q'(t)$
for $t\in[\tau(q)/2,\tau(q)]$ and the length $\tau(q)/2$ of
$\gamma_q\big|_{[\tau(q)/2,\tau(q)]}$.  The same Hessian bound will hold for
$\phi_p$.

Now let $K\subset M$ be compact.  Then by Proposition~\ref{speed-bd} the
vectors $\gamma_q'(\tau(q))$ for $q\in K$ are all contained in some compact
set $\widehat{K}$ of the tangent bundle of $M$.  Therefore there is a
compact set $K_1\subset M$ that will contain all the segments
$\gamma_q\big|_{[\tau(q)/2,\tau(q)]}$ with $q\in K$ and a compact set
$\widehat{K}_1\subset T(M)$ that will contain all the tangent vectors to
these segments.
Therefore there are uniform upper and lower bounds for
both the sectional curvatures of two planes containing a tangent vector to
all of the segments $\gamma_q\big|_{[\tau(q)/2,\tau(q)]}$ and also the
lengths $\tau(q)/2$ of these segments.
It follows that there are uniform
two sided bounds on the Hessians for the support functions $\phi_q$ for
$q\in K$.
Therefore given any point $q_0$ and a compact
coordinate neighborhood $K$ of $q_0$, by writing out the the two sided Hessian
bounds in terms of the coordinates we find that the lower support functions
$\phi_q$, $q \in K$, to $\tau$ will also satisfy two sided bounds on the Hessian
$D^2\phi_q(q)$ with respect to the coordinates.  Therefore
Lemma~\ref{quad-spt} implies $\tau$ is semiconvex near $q$.  As $q$ was any
point of $M$ this completes the proof.~\end{proof}

We now consider further the regularity of the level sets
$S_a:=\{q:\tau(q)=a\}$ of
the cosmological time function.  To do this it is convenient to work in some
special coordinate systems.  Let $q$ be any point of $M$ and let $N_0$ be a
smooth spacelike hypersurface passing through $q$. Let $(x^1\cd x^{n-1})$
be local coordinates on $N_0$ centered at $q$ and let $x^n$ be the
signed Lorentzian distance (with $x^n$ positive to the future of $N_0$ and
negative to the past).  Then near $q$, $(x^1\cd x^n)$ is a local coordinate
system so that the form of the metric in this coordinate
system is
$$
g=\sum_{A,B=1}^ng_{AB}dx^Adx^B=\sum_{i,j=1}^{n-1}g_{ij}dx^idx^j-(dx^n)^2.
$$
Call such a coordinate system an {\bi adapted coordinate system centered at
$q$}.  Then
for any spacelike hypersurface $N$ of $M$ through $q$ we have that
locally $N$ can be parameterized as the graph of a function $f$.  That
is,
\begin{equation}\label{para}
F_f(x^1\cd x^{n-1})=(x^1\cd x^{n-1},f(x^1\cd x^{n-1})).
\end{equation}

\begin{prop}\label{levelsets}
Let the cosmological time function $\tau$ of $(M,g)$ be regular and for
$a\in (0,\infty)$ let $S_a=\{x:\tau(x)=a\}$ be a nonempty level set of
$\tau$.  Then
for any $q\in S_a$ and every adapted coordinate system $x^1\cd x^n$
centered at $q$ there is a local parameterization of $S_a$ of the
form~(\ref{para}) for a unique function~$f$ defined on an neighborhood of
the origin in $\r^{n-1}$.  This function~$f$ is semiconcave (that is $-f$
is semiconvex) and therefore it is locally Lipschitz and its first and second
derivatives exist almost everywhere.
\end{prop}

\begin{proof}
The existence and uniqueness of the function~$f$ is elementary, and follows
from the fact
that $S_a$ is an acausal hypersurface.  We now are
going to construct upper support functions for $S_a$ at each of its points.
For any $p\in S_a$ let $\gamma_p:(0,\tau(p)]\to M$ be a ray that realizes
the distance to the initial singularity of $M$ in the sense of
Proposition~\ref{prop:rays}.  Then define
$$
\Sigma_p:=\{x\in I^+(\gamma_p(a/2)):d(\gamma_p(a/2),x)=a/2\}.
$$
That is, $\Sigma_p$ is the future Lorentzian distance sphere of radius
$a/2$ about the point $\gamma_p(a/2)$.  Using that $\gamma_p$ realizes the
distance between any two its points and that $\gamma_p(\tau(a))=p$ we see
$d(\gamma_p(a/2),p)=d(\gamma_p(a/2),\gamma_p(a))=a/2$ so that $p$ is
in $\Sigma_p$.  Also using the reverse Lipschitz inequality for $\tau$, if
$x\in \Sigma_p$ then
$$
\tau(x)\ge \tau(\gamma_p(a/2))+d(\gamma_p(a/2),x)=\frac{a}2+\frac{a}2=a.
$$
Thus every point of $\Sigma_p$ is in the causal future of $S_a$.  As $\gamma_p$
is maximizing, the segment $\gamma_p\big|_{[a/2,a]}$ will be free of
conjugate points and therefore the $\Sigma_p$ is a smooth hypersurface in a
neighborhood of $p$.  Now let $K\subset S_a$ be a compact set.  Then by
Proposition~\ref{prop:rays} the set $\{\gamma_p'(a):p\in K\}$ has compact
closure in $T(M)$.  Therefore an argument like that used in the proof of
Proposition~\ref{semiconvex} (based on elementary comparison theory)
implies that if $h^{\Sigma_p}_p$ is the second fundamental form of
$\Sigma_p$ at the point $p$ then $h^{\Sigma_p}_p$ satisfies a uniform two
sided bound for $p\in K$ (or, what is the same thing, the absolute values of
the principle curvatures of $\Sigma_p$ at the point $p$ are uniformly
bounded for $p\in K$).

For $p\in S_a$ sufficiently close to $q$ we can parameterize $\Sigma_p$ by a
function $F_{f_p}$ with $F_{f_p}$ defined as in~(\ref{para}).  As the
hypersurfaces $\Sigma_p$ are in the causal future of $S_a$  the functions
$f_p$ satisfy $f_p\ge f$ near~$p$ and thus they are upper support
functions for $f$ near $p$.
The bound  on the second fundamental forms
of the $\Sigma_p$'s can be translated into a bound on the Hessians $D^2f_p$
(for the details of this calculation see~\cite{AGH-max}).  Therefore
Lemma~\ref{quad-spt} implies $-f$ is semiconvex.  This completes the
proof.~\end{proof}

\section{Examples}

\subsection{A globally hyperbolic spacetime with $\tau$
finite valued but discontinuous}
Let $\phi:\r\to[0,\infty)$ be a smooth function with support in the
interval $[1/4,1/2]$ and with
$\int_{-\infty}^\infty\phi(t)\,dt=\int_{1/4}^{1/2}\phi(t)\,dt=2$. Define a
function $\Phi$ on the upper half plane $M:=\{(x,y):y>0\}$ by
$$
\Phi(x,y)=\left\{\begin{array}{rcl} 1+\dfrac1x\phi\(\dfrac{y}{x}\),&& x>0\\
                &&\\    1,&& x\le0.\end{array}\right.
$$
Let $g$ be the Lorentzian metric on $M$ given by
$$
g:=dx^2-\Phi(x,y)^2\,dy^2.
$$

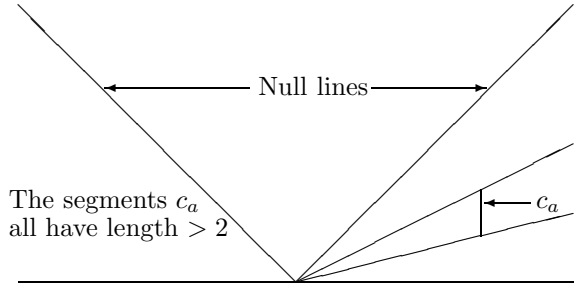
\begin{figure}[ht]
\begin{center}\unitlength=.7pt
\begin{picture}(300,150)

\put(0,0){\line(1,0){300}}
\put(150,0){\line(2,1){150}}
\put(150,0){\line(4,1){150}}
\put(130,101){\small Null lines}
\put(125,105){\vector(-1,0){78}}
\put(191,105){\vector(1,0){62}}
\put(250,25){\line(0,1){25}}
\put(280,40){\small$c_a$}
\put(277,43){\vector(-1,0){25}}

\put(-5,40){\small The segments $c_a$}
\put(-5,25){\small all have length $>2$}

\thinlines
\put(150,0){\line(-1,1){150}}
\put(150,0){\line(1,1){150}}

\end{picture}
\end{center}
\end{figure}

\noindent
Then this metric is smooth on $M$ and using $\int_{1/4}^{1/2}\phi(t)\,dt=2$
it is not hard to check that for any $a>0$ the length of the timelike curve
$c_a:[a/4,a/2]\to M$ given by $c_a(t):=(a, t)$ has Lorentzian length
$L(c_a)=\int_{a/4}^{a/2}\Phi(a,t)\,dt=2+a/4$. Let $g_0=dx^2-dy^2$ be the
standard flat Lorentzian metric on $M$ and let $W$ be the open wedge
$W:=\{(x,y): x>0, y>0, x/4< y < x/2\}$.  Then $g=g_0$ outside of $W$.  If
$F:=I^+(W)\setminus W=\{(x,y):y>0, -y< x\le y/2\}$ then, using that the
segments $c_a$ all have Lorentzian length greater than $2$, we see that
$$
\tau(x,y)>2\quad\mbox{for all}\quad (x,y)\in F.
$$
But for $(x,y)\not\in I^+(W)$ the existence time of $(x,y)$ is the distance
of $(x,y)$ from the $x$-axis in the usual metric $g_0$, that is
$$
\tau(x,y)= y\quad\mbox{for all}\quad (x,y)\in M\setminus I^+(W).
$$
This implies that $\tau$ is discontinuous at each point of the segment
$\{(x,y): -2<x<0, y=-x\}$.
But the spacetime $(M,g)$ is globally hyperbolic
and has finite existence times.  \smallskip

\subsection{Non-strongly causal spacetimes with $\tau$ finite valued}
Consider the well-known example of a spacetime which is causal
but not strongly causal
(cf.~\cite[p~193~Figure~38]{Hawking-Ellis}).  In this example, which
is a cylinder with slits, it is easily verified that $\tau$ is
finite valued.

If we are willing to drop the requirement that the metric of $(M,g)$ is
smooth, but only of class $C^1$ then there is an example of a Lorentzian
metric on a cylinder that has $\tau$ finite valued, but which has a closed
causal curve (which turns out to be a null geodesic).
This example, which we now describe, is used in the next subsection
to construct a spacetime with a nonregular $\tau$ such that
$\tau \to 0$ along all past inextendible timelike geodesics.

Let the circle $S^1$
(which we think of as $\r$ modulo $2\pi$)
have coordinate $x$ and for any $\alpha>0$ define a metric on the
space $M:=S^1\times \r$ by
$$
g:=dx dt +|t|^{2\alpha}\,dx^2=dx(dt+|t|^{2\alpha}\,dx).
$$
At each point the null directions are defined by $dx=0$ and
$dt+|t|^{2\alpha}dx=0$.  If the direction of $\f/\f t$ is used as the
direction of increasing time then the only closed causal curve is the curve
$\{t=0\}$.
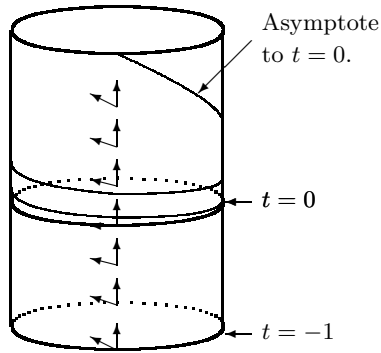
\begin{figure}[ht]
\begin{center}
\begin{picture}(100,135)

\put(101,60){\vector(-1,0){10}}
\put(105,58){\footnotesize $t=0$}

\put(101,60){\vector(-1,0){10}}
\put(105,58){\footnotesize $t=0$}

\put(101,10){\vector(-1,0){10}}
\put(105,8){\footnotesize $t=-1$}

\put(105,125){\footnotesize  Asymptote}
\put(105,113){\footnotesize to $t=0$.}
\put(101,121){\vector(-1,-1){20}}

\put(50,51){\vector(0,1){10}}
\put(50,51){\vector(-1,0){10}}

\put(50,66){\vector(-4,1){10}}
\put(50,66){\vector(0,1){10}}

\put(50,81){\vector(-3,1){10}}
\put(50,81){\vector(0,1){10}}

\put(50,96){\vector(-2,1){10}}
\put(50,96){\vector(0,1){10}}

\put(50,36){\vector(-4,1){10}}
\put(50,36){\vector(0,1){10}}

\put(50,21){\vector(-3,1){10}}
\put(50,21){\vector(0,1){10}}

\put(50,4){\vector(-2,1){10}}
\put(50,4){\vector(0,1){10}}

\thicklines

%Top oval
\bezier{147}(10,125)(10,133)(50,134)
\bezier{147}(90,125)(90,133)(50,134)
\bezier{147}(10,125)(10,117)(50,116)
\bezier{147}(90,125)(90,117)(50,116)

%Middle oval
\bezier{15}(10,60)(10,68)(50,69)
\bezier{15}(90,60)(90,68)(50,69)
\bezier{147}(10,60)(10,52)(50,51)
\bezier{147}(90,60)(90,52)(50,51)

%Bottom oval
\bezier{15}(10,13)(10,21)(50,22)
\bezier{15}(90,13)(90,21)(50,22)
\bezier{147}(10,13)(10,5)(50,4)
\bezier{147}(90,13)(90,5)(50,4)

\put(10,13){\line(0,1){112}}
\put(90,13){\line(0,1){112}}

%The spiral

\thinlines

\bezier{147}(90,61)(90,54)(50,54)
\bezier{147}(50,54)(10,56)(10,65)

\bezier{147}(90,69)(90,63)(50,63)
\bezier{147}(50,63)(10,66)(10,75)

\bezier{147}(90,90)(90,98)(50,116)

\end{picture}
\end{center}
\end{figure}

A past inextendible causal curve will either diverge along
the cylinder to $t=-\infty$ or be asymptotic to the the null geodesic
$\{t=0\}$ as in the figure.  We now show that any past inextendible
causal curve asymptotic to $\{t=0\}$ starting at $(x_0,t_0)$ has length
bounded just in terms of $(x_0,t_0)$.  In doing this it is convenient
to work on the universal cover of the cylinder, that is $\r^2$. And
in doing the preliminary part of the calculation it is no harder to work
with a slightly more general class of metrics.
Let $f(x)$ be any smooth positive function
defined on the real line (in our example  $f(x)\equiv 1$) and let
$\phi(t)$ be a $C^1$ function so that $t=0$ is the only zero of $\phi$
(in our example $\phi(t)=|t|^\alpha$) again defined on the real line.
As $t=0$ is the only zero of $\phi$ it does not change sign on $(0,\infty)$
and we assume that $\phi(t)>0$ on $(0,\infty)$.
Define a Lorentzian metric on $\r^2$ by
$$
g_0=dxdt+\phi(t)^2f(x)^2\,dx^2=dx(dt+\phi(t)^2f(x)^2\,dx)
$$
and use the time orientation so that $\f /\f t$ points to the future.
At each point the null directions are defined by $dx=0$ and
$dt+\phi(t)^2f(x)^2\,dx=0$. From this it follows  that $\{t=0\}$ is a null
geodesic and that every past inextendible causal curve $c$ either is
divergent with
$t\to -\infty$ along $c$ or $c$ remains in the closed upper half plane
defined by $t\ge 0$ and $c$ is asymptotic to the null geodesic $\{t=0\}$ in
such a way that $x$ is monotone increasing along $c$.

Now let $c$ be a past inextendible causal curve starting at the point
$(x_0,t_0)$ and so that $c$ is asymptotic to the null geodesic $\{t=0\}$.
Then $c$ has a parameterization of the form $c(t)=(x(t),t)$ defined on
$(0,t_0]$.  As this curve is causal we have (using the notation
$\dot{x}=dx/dt$),
\begin{align*}
0&\ge g_0(c'(t),c'(t))= \dot{x}+\phi(t)^2f(x)^2\dot{x}^2\\
&=\(\dot{x}\phi(t)f(x)+\frac{1}{2\phi(t)f(x)}\)^2-\frac{1}{4\phi(t)^2f(x)^2}\\
&\ge \frac{-1}{4\phi(t)^2f(x)^2},
\end{align*}
%
%
%$$
%\dot{x}+\phi(t)^2f(x)^2\dot{x}^2=\dot{x}(1+\phi(t)^2f(x)^2\dot{x})\le 0.
%$$
%This implies $-1/(\phi(t)^2f(x)^2)\le \dot{x}\le 0$ which implies
%
%(The second of these estimate follows by noting the lowest point on the
%function $\dot{x}\mapsto \dot{x}+\phi(t)f(x)\dot{x}^2$ occurs when
%$\dot{x}=-1/(2\phi(t)f(x)\dot{x}^2)$.)
and thus,
\begin{equation}\label{dot-bds}
        |\dot{x}+\phi(t)^2f(x)^2\dot{x}^2|\le \frac1{4\phi(t)^2f(x)^2}.
\end{equation}
As $c$ is asymptotic to $\{t=0\}$ it
follows that $t\ge 0$ and thus also $\phi(t)\ge 0$ along $c$.  Thus the
Lorentzian length of $c$ satisfies
$$
L(c)=\int_0^{t_0}\sqrt{\dot{x}+\phi(t)^2f(x)^2\dot{x}^2}\,dt
\le \int_0^{t_0}\frac{dt}{2\phi(t)f(x(t))}
$$
where the inequality follows from using the bound in~(\ref{dot-bds}).
Now letting $\phi(t)=|t|^\alpha$ with $0<\alpha<1$ and $f(x)\equiv 1$
then this leads to the bound $L(c)\le t_0^{1-\alpha}/(2(1-\alpha))$ as
required.

Now if we let $M:=\{(x,t)\in S^1\times \r: t> -1\}$ then the bound on
the length of curves asymptotic to $\{t=0\}$ just given implies that if
$0<\alpha <1$ and $M$ has the metric $g=dxdt+|t|^{2\alpha}\,dx^2$ then
$(M,g)$ has $\tau$ finite valued, but $\tau$ does not go to zero along
the inextendible causal curves asymptotic to $\{t=0\}$.  It is
worth noting that in this example $\tau$ is continuous.

We know of no example where $\tau$ is finite, there are closed causal
curves, and the metric is smooth.

\subsection{A non-regular $\tau$ going to zero along all past inextendible
causal
geodesics}\label{sec:null-example}

The definition of $\tau$ being regular requires that $\tau$ go to zero
along all past inextendible causal curves.  It is natural to ask if this can be
weakened to only requiring that $\tau$ go to zero along all past inextendible
causal geodesics.  Here we give an example to show that this is not the
case.  Like the example just given the metric in this example is of
class $C^1$ but not $C^2$.

First let $(M_2,g_2)=(S^1\times (-1,\infty), dxdt+|t|^{2\alpha}dx^2)$
be the two dimensional example just given (so that $0<\alpha<1$) and set
$$
f(y)=e^{y^2}-1.
$$
Note that $f(0)=0$ and $f(y)>0$ for $y\ne0$.
Let $M:=\{ (x,y,t)\in
S^1\times\r\times \r: t>-1\}$ with the metric
$$
g:=dy^2+e^{2y}(dxdt +(|t|^{2\alpha}+f(y))dx^2).
$$
Then the two dimensional submanifold defined by $y=0$ is isometric to
$(M_2,g_2)$.  Moreover this submanifold is totally umbilic in $(M,g)$
and so no curve in $(M_2,g_2)$ can be a geodesic in $(M,g)$.  Let
$\eta$ be the null geodesic defined by $\{t=0, y=0\}$.  The following
is easy to verify.

\begin{lemma} Let $c$ be a past inextendible causal curve in $(M,g)$. Then
one of the following holds:
\begin{enumerate}
\item $t\to -1$ along $c$ and $c$ runs off of the ``bottom'' of $M$ (that
is the part of the boundary defined by $t=-1$).
\item $t\to 0$ along $c$ and $c$ is asymptotic to the closed null
curve $\eta$.  Neither $\eta$ or any curve asymptotic to it are
geodesics.~\qed
\end{enumerate}
\end{lemma}

Harder to show is:

\begin{lemma}
Let $c$ be a past inextendible curve starting at the point
$(x_0,y_0,t_0)$ which is asymptotic to the null curve $\eta$.  Then
there is a finite upper bound on the length of $c$ only depending on
$t_0$.
\end{lemma}

\begin{proof}
Analogous to what was done in the last example, there is a
parameterization of $c$ of the form $c(t)=(x(t),y(t),t)$ with $t\in
(0,t_0]$.  As $c$ is causal $g(c'(t),c'(t))\le 0$ which implies,
\begin{align*}
0&\ge g(c'(t),c'(t))=\dot{y}^2+e^{2y}\(\dot{x}+(|t|^{2\alpha}+f(y))\dot{x}^2\)\\
&=\dot{y}^2+e^{2y}\(\frac{1}{2\sqrt{|t|^{2\alpha}+f(y)}}+\sqrt{|t|^{2\alpha}
+f(y)}\,\dot{x}\)^2
        -\frac{e^{2y}}{4(|t|^{2\alpha}+f(y))}\\
&\ge -\frac{e^{2y}}{4(|t|^{2\alpha}+f(y))}
\end{align*}
and thus,
$$
\sqrt{|g(c'(t),c'(t))|}\le \frac{e^y}{2\sqrt{|t|^{2\alpha}+f(y)}}.
$$
If $y\le 3$ then,
$$
\frac{e^y}{2\sqrt{|t|^{2\alpha}+f(y)}} \le
\frac{e^3}{2\sqrt{|t|^{2\alpha}}}=\frac{e^3}{2|t|^{\alpha}}\le
\frac{12}{|t|^\alpha}.
$$
If $y\ge 3$ then $e^y\le \sqrt{e^{y^2}-1}=\sqrt{f(y)}$ and so
$$
\frac{e^y}{2\sqrt{|t|^{2\alpha}+f(y)}}\le \frac{e^y}{2\sqrt{f(y)}} \le
\frac{1}{2}.
$$
Putting these together we have,
$$
\sqrt{|g(c'(t),c'(t))|} \le \max\( \frac{12}{|t|^\alpha}, \frac{1}{2}\),
$$
which implies,
$$
L(c)=\int_0^{t_0}\sqrt{|g(c'(t),c'(t))|} \,dt\le
\int_0^{t_0} \max\( \frac{12}{|t|^\alpha}, \frac{1}{2}\)\,dt,
$$
which is finite as $0<\alpha<1$.  This gives the required bound and
completes the proof of the lemma.~\end{proof}

Therefore in the spacetime $(M,g)$ all past inextendible curves $c$
either have $t\to -1$ along $c$ (in which case $\tau\to 0$ along $c$)
or $c$ is asymptotic to the null curve $\eta$ (in which case $\tau$
does not go to zero).  As no geodesics are asymptotic to $\eta$ this
gives the an example of a spacetime where $\tau\to 0$ along all
inextendible causal geodesics, but which is not regular.  It would be
interesting to know if there is a smooth example where this happens.
\bigskip

{\small {\bf Acknowledgments.}  Joe Fu pointed out the relevance of the
reference~\cite{Bangert:gen-convex} to the results of
Section~\ref{sec:level}.  Some comments of Jeeva Anandan were also
useful in writing the paper.}

%bibliographystyle{amsplain}
%bibliography{/math/faculty/howard/tex/inputs/HowRefs}
%\bibliography{tj:Refs/HowRefs,cosmo-refs}
%\begin{thebibliography}{1}

\ifx\undefined\bysame
\newcommand{\bysame}{\leavevmode\hbox to3em{\hrulefill}\,}
\fi

\end{document}